\documentclass{ifacconf}
\usepackage{graphicx}      
\usepackage{natbib}        
\usepackage{amsmath,amssymb,amsfonts}

\usepackage{algorithmicx}
\usepackage[noend]{algpseudocode}

\usepackage{graphicx}
\usepackage{textcomp}
\usepackage{xcolor}
\usepackage{subcaption}
\usepackage{caption}
\usepackage{mathabx}
\usepackage{algorithm} 
\usepackage{makecell} 
\usepackage{enumitem}
\usepackage{url}

\newtheorem{remark}{Remark}
\DeclareMathOperator*{\argmin}{argmin}
\DeclareMathOperator*{\argmax}{argmax}

\newcommand{\norm}[1]{\left\lVert#1\right\rVert}

\newcommand{\sdbcheck}[1]{\textcolor{black}{#1}}

\usepackage{mathtools}
\DeclarePairedDelimiter{\ceil}{\lceil}{\rceil}

\begin{document}
\begin{frontmatter}

\title{Vector Cost Bimatrix Games with Applications to Autonomous Racing\thanksref{footnoteinfo}} 

\thanks[footnoteinfo]{This work was supported by the National Science Foundation CAREER Award ECCS-2236537. Special thanks also to the Spartan Autonomous Racing Club at Michigan State.}

\author[First]{Benjamin R. Toaz} \qquad 
\author[First]{Shaunak D. Bopardikar} 

\address[First]{Michigan State University, 
   East Lansing, MI 48824 USA \\ (E-mails: toazbenj@msu.edu, shaunak@egr.msu.edu)}

\begin{abstract}                
We formulate a vector cost alternative to the scalarization method for weighting and combining multi-objective costs. The algorithm produces solutions to bimatrix games that are simultaneously pure, unique Nash equilibria and Pareto optimal with \sdbcheck{guarantees for avoiding worst case outcomes}. We achieve this by enforcing exact potential game constraints to guide cost adjustments towards equilibrium, while minimizing the deviation from the original cost structure. The magnitude of this adjustment serves as a metric for differentiating between Pareto optimal solutions. \sdbcheck{We implement this approach in a racing competition between agents with heterogeneous cost structures, resulting in fewer collision incidents with a minimal decrease in performance. Code is available at} \url{https://github.com/toazbenj/race_simulation}.
\end{abstract}


\end{frontmatter}

\section{Introduction}

Problems in inverse game theory involve the structuring of incentives in order to produce a desired behavior. Mechanism design involves designing a cost structure such that autonomous agents pick actions according to their own self-interest that also align with the larger goals laid out by the designer. When applying this framework to autonomous vehicle decision making, we immediately see the dichotomy between maximizing performance in the form of efficient trajectory planning and the desire to avoid unsafe situations.

One of the most common ways to structure noncooperative games is using the zero-sum structure that pits two players against each other in the pursuit of a single objective. During trajectory planning, this cost structure works well for simple competitive interactions (\cite{StochasticDynamic}) and has been extensively modified for adversarial training of agents using machine learning techniques (\cite{AUnifiedGame}), (\cite{TowardsSafer}). 

Collapsing several types of costs into composites of outcomes using zero-sum games as a base structure has allowed for the moderation of competitive behavior with the need to observe safety constraints (\cite{EncodingDefensive})(\cite{GameTheoreticPlanning})(\cite{StochasticGames}). Games that are formulated in this way often take on both cooperative and noncooperative aspects (\cite{ANoncooperativeGame}). Using potential games to find admissible Nash equilibria for semi-cooperative interactions has also found some success, though both of these objectives still are collapsed into a single scalar cost using weighted sums (\cite{PotentialGameBased}). Extensions of potential games with approximated Nash equilibria for multi-agent racing have also been implemented, while objective functions are limited to progress relative to a reference trajectory, with safety or other goals encoded into separate optimization constraints (\cite{AlphaRacer}). Weights can be tuned using methods such as inverse reinforcement learning (\cite{AlgorithmsForInverse}), optimization algorithms (\cite{GameTheoreticMotion}), or learning user preferences (\cite{ToLeadOr}).

Some works have explored how to tune weights in order to achieve a better balance between outcomes in multi-objective problems, especially using techniques in reinforcement learning (\cite{ReinforcementLearningAnd})(\cite{AlgorithmsForInverse}). These have had some success in creating behaviors that approach the optimum, but do not have safety guarantees with respect to worst case outcomes (\cite{DeepLatent}). Altering the structure of weight optimization algorithms has also allowed for additional trade offs to be considered by changing the topography of the Pareto frontier (\cite{ScalarizingMultiObjective}). Although this \emph{scalarization} approach is simple to implement, it does not guarantee any sort of optimality with respect to the original vector of costs, either in the sense of the Pareto optimality or in that of a Nash equilibrium. In fact, inadequately tuned weights may correspond with worst-case choices for the unweighted starting costs. 

The use of set theory and approachability has had some success at reasoning how to avoid or force a given outcome within a vector cost space, which can be applied in the creation of safety guarantees (\cite{AnAnalogOf}). Vector cost approaches generally involve finding Pareto optimal solutions that are equivalent to all other outcomes on the Pareto frontier (\cite{GamesWithVector}). Once the Pareto optimal solutions are established, policies can be selected by choosing a desired trade-off between different types of outcomes, even changing in the middle of a game (\cite{BridgingTheGap}). 

Within the literature, there is a deficit of approaches that allow for efficient evaluation of Pareto optimal policies. Quantifying the desirability of any given section of the Pareto front remains a challenge, and this work contributes a benchmark to select from these outcomes. Structuring our bimatrix game using vector costs allows us to circumvent some of the shortcomings faced by these previous works while solving multi-objective problems.

In this paper, we present an alternative to scalarization using methods inspired by mechanism design (Section~\ref{sec:problem}). Our major contributions detailed in Section~\ref{sec:approach} include the formulation of a convex optimization algorithm. It produces solutions to the vector-cost bimatrix game by adjusting the costs of the ego player, while leaving the nonego player with fixed scalarized costs. Our approach is based on imposing a potential structure on the costs. The resulting policies have the following desirable properties: (1) Pareto optimality with respect to the original vector costs, (2) a unique, \emph{pure} Nash equilibrium, (3) matching the security policies of both players, and (4) a guarantee on avoiding worst-case scenarios. An additional byproduct of this approach is that the amount of adjustment needed to achieve costs that satisfy these properties can be used as a metric to distinguish between seemingly interchangeable points on the Pareto frontier.  

We demonstrate the application of our algorithm as an online trajectory planning method within a racing game between autonomous vehicles in Section~\ref{sec:application}. The passing game involves the attacking vehicle attempting to pass the slower defending vehicle while staying on the track and avoiding collisions. \sdbcheck{We found that our approach exhibited significantly fewer collisions between vehicles, with the trade-off of a small drop in competitive performance.}

\section{Problem Formulation}\label{sec:problem}

A vector cost bimatrix game consists of 2 players, both of whom are minimizers. Each player has a finite number of actions that they choose out of respective actions sets $U_1:=\{1,\dots, m\}$ and $U_2 = \{1,\dots, n\}$, and the players choose these actions simultaneously. Their policies are denoted by $\gamma \in U_1$ for player 1 and $\sigma \in U_2$ for player 2. 
Here the information structure is full information, simultaneous play in the space of pure policies. For ease of presentation, we consider only two costs in this paper. The outcome vectors for the game are $J_i : U_1 \times U_2 \to \mathbb{R}^{2}, \text{ for } i \in \{1,2\},$ that define the costs for two objectives given by two given matrices $A_i,B_i \in \mathbb{R}^{n \times m}$, as per $J_i(\gamma,\sigma)= (A_i(\gamma,\sigma), B_i(\gamma,\sigma))$.

The first cost defined by $A_1$ and $A_2$ adheres to a zero sum game structure, meaning $A_1 = -A_2$. This is meant to encode the competitive aspects of the game where each player aims to perform better than the other. The second cost defined by $B_1$ and $B_2$ is assumed to admit a potential function $\phi \in \mathbb{R}^{n \times m}$ which satisfies the following property. For every pair of policies $\gamma, \sigma \in U_1\times U_2$, 
\begin{equation}\label{eq:exact potential}
    \begin{aligned}
        B_1(\gamma,\sigma)-B_1(\widebar{\gamma},\sigma) =
        \phi(\gamma,\sigma)-\phi(\widebar{\gamma},\sigma), \forall \bar{\gamma} \in U_1, \\
        B_2(\gamma,\sigma)-B_2(\gamma,\widebar{\sigma}) =
        \phi(\gamma,\sigma)-\phi(\gamma,\widebar{\sigma}), \forall \bar{\sigma} \in U_2.
\end{aligned}
\end{equation}

An example of this cost structure arises in an \emph{identical interests game} where $B_1 = B_2$. In this case, the costs $B_1$ and $B_2$ encapsulate the shared goals of the players to adhere to certain constraints. \sdbcheck{For an autonomous vehicle, this could come in the form of the needing to stay on the road and avoid collisions with other vehicles, something that we loosely term "safety."} For potential games, any global minimum of $\phi$ matches an admissible Nash equilibrium $\{\gamma^*,\sigma^*\}$ of the associated bimatrix game defined by $B_1$ and $B_2$ in which the policy pair $\{\gamma^*,\sigma^*\}$ satisfies
\begin{equation}
B_1(\gamma^*, \sigma^*) \leq 
B_1(\gamma,\sigma^*), \, 
B_2(\gamma^*, \sigma^*) \leq 
B_2(\gamma^*, \sigma),
\label{equilibrium} \end{equation}
for every $\gamma \in U_1$ and for every $\sigma \in U_2$.

The scalarization approach (e.g., (\cite{ScalarizingMultiObjective})), simplified for a two cost application, can be represented as picking a set of weights $\theta_1 \in \mathbb{R}, \theta_2 \in \mathbb{R}$, and combining the two vector costs into composite matrices as
\begin{equation}
    C_i = \theta_1A_i+\theta_2B_i, \forall i \in \{1,2\}.
\label{scalarization}\end{equation}
In the context of the bimatrix game defined by matrices $C_1$ and  $C_2$, each player can now play a security policy $\{\gamma^s,\sigma^s\}$ for this bimatrix game, and achieve outcomes that avoid worst-case scenarios in the space of combined weighted costs. Specifically, 
\begin{equation}\label{eq:security}
    \gamma^{s} \in \argmin_{\gamma} \max_{\sigma} C_1(\gamma, \sigma),\quad \sigma^{s} \in \argmin_{\sigma} \max_{\gamma} C_2(\gamma, \sigma).
\end{equation}
A security policy pair $\{\gamma^s, \sigma^s \}$ may constitute a Nash equilibrium if it satisfies the conditions described in Equation~\eqref{equilibrium}. If each player plays their security policy, then we term the outcome as the solution to the scalar cost bimatrix game defined by $C_1$ and $C_2$.

The objective of the vector cost game is to select policies $(\gamma, \sigma)$ that minimize \emph{both} costs in $J_i$ for both players $i\in \{1,2\}$. To account for the multi-objective nature of the problem, we define a set of optimal policies in vector cost space using \emph{Pareto optimality}. Given a fixed opponent policy $\sigma$, policy $\gamma^*$ is within the set of Pareto optimal policies $\mathcal{P}_1$ with respect to the cost $J_1$ if at least one component of $J_1(\gamma^*,\sigma)$ dominates all other outcomes and the other costs are no worse. Formally, this is given by
\begin{multline}\label{eq:pareto}
    \mathcal{P}_1(A_1, B_1, \sigma) = \Big \{\gamma^* \in U_1 \, : \, A_1(\gamma^*,\sigma) \leq A_1(\gamma,\sigma) \text{ and } \\ B_1(\gamma^*,\sigma) \leq B_1(\gamma,\sigma), \forall \gamma \in U_1\Big \}.
\end{multline}

We refine $\mathcal{P}_1$ further into a smaller subset that also has maximum cost guarantees. A worst-case policy player 1 for any of the cost types is $\gamma^w$, which lies within the set of policies $\mathcal{W}_1$ defined as
\begin{multline}\label{eq:worst}
\mathcal{W}_1(A_1, B_1, \sigma) = \argmax_{\gamma^w \in U_1}A_1(\gamma,\sigma) \bigcup \argmax_{\gamma^w \in U_1}B_1(\gamma,\sigma).
\end{multline}
For example, this outcome may occur if player 1 simply plays $\gamma^{s}$ according to competitive costs $A_1$, neglecting $B_1$, or vice versa. These outcomes would be corner points at either extreme of the Pareto frontier defined by $\mathcal{P}_1$. Assuming that player 2 selects policy $\sigma^s$ as per Equation~\eqref{eq:security}, for a given pair $\theta_1$ and $\theta_2$, we are concerned with the set of policies $\{\gamma^m,\sigma^s\}$ where $\gamma^m \in \mathcal{M}_1$ defined as 
\begin{multline}\label{eq:moderate}
      \mathcal{M}_1(A_1, B_1, \sigma) = \Big \{\gamma^m \in U_1\, : \, \gamma^m \in \mathcal{P}_1(A_1, B_1, \sigma) \textbf{ and }
      \\ \gamma^m \notin \mathcal{W}_1(A_1, B_1, \sigma) \Big \}.
\end{multline}
The set $\mathcal{M}_1$ can be considered as the set of \emph{moderate} Pareto optimal policies for the game with vector costs.

\noindent{\bf Problem statement:} For an arbitrary value of $\theta_1$ and $\theta_2$ known to both players, suppose that player 2 uses a security policy $\sigma^s$ defined in Equation~\eqref{eq:security}. Define an error matrix $E \in \mathbb{R}^{n \times m}$ such that the adjusted costs for player 1 denoted by $\tilde{A_1}$ are produced by the sum $\tilde{A_1} = E+A_1$. The goal is to find the smallest error matrix $E$ in the Frobenius norm for the bimatrix game defined by the pair $\tilde{A_1},C_2$ such that $\tilde{A}_1$ yields a unique security policy $\tilde{\gamma}^s$ where $\tilde{\gamma}^s\in \mathcal{M}_1(A_1, B_1, \sigma^s)$. 

\begin{remark}[{\bf Cost Structure}]
     This approach's main utility is to allow a vector cost player to select policies that lead to balanced outcomes with respect to its own goals. We use the scalarization method as a baseline to model the decisions of player 2 while player 1 uses multiple objectives to make decisions. When player 1 is presented with a number of possible policies, some of these will be Pareto optimal in the sense of Equation~\eqref{eq:pareto}. These may cause extreme outcomes that maximize one cost while minimizing the other, such as in Equation~\eqref{eq:worst}.
     
     To minimize all component costs while guaranteeing avoidance of worse-case outcomes, we transform the competitive costs to be more similar to the safety costs while preserving portions of their original structure, which is difficult to accomplish using scalarization. We do this so that the adjusted costs will yield a security policy that is both Pareto optimal and safe in the space of the original costs, as per Equation~\eqref{eq:moderate}. 
\end{remark}

\section{Proposed Approach and Main Results}\label{sec:approach}

In this section, we introduce our cost adjustment algorithm that produces a new cost matrix for player 1 to yield security policies that have both Pareto optimality and outcome guarantees. We propose to achieve this through the solution of a convex optimization problem that minimizes the error $E$ while adhering to constraints informed by potential functions. In order for the optimization to be feasible, the fixed costs of the opponent (player 2) must have additional properties, which we elaborate on below.    


In order to inform our choice of $E$ that will define $\tilde{A_1}$, we aim to produce a potential function $\phi \in \mathbb{R}^{n \times m}$ from the bimatrix pair $\tilde{A}_1, C_2$ that adheres to the exact potential game structure according to Equation~\eqref{eq:exact potential}. We impose that the minimum entries of $\phi$ match with the Nash equilibria of the corresponding bimatrix game. When there is one unique global minimum of $\phi$, we would like the unique admissible Nash equilibrium to be equivalent to the pair of security policies $\{\tilde{\gamma}^s,\sigma^s\}$ for that same bimatrix game. 

This leads us to the following optimization problem. The input is the desired location of the minimum of $\phi$ given by entry $(r,c)$ and the cost matrices $A_1$ and $C_2$.  
\begin{equation}
\begin{aligned}
& \min_{E, \phi \,: \, \phi(r,c) = 0} && \|E\|_F^2\\[1ex]
& \text{subject to} && \forall \gamma, \bar{\gamma} \in U_1, \text{ and } \forall \sigma, \bar{\sigma} \in U_2, 
\\& && A_1(\gamma,\sigma)+E(\gamma,\sigma)-A_1(\widebar{\gamma},\sigma)-E(\widebar{\gamma},\sigma) =\\
& && \quad  \phi(\gamma,\sigma)-\phi(\widebar{\gamma},\sigma), \quad \\
& && C_2(\gamma,\sigma)-C_2(\gamma,\widebar{\sigma}) =\phi(\gamma,\sigma)-\phi(\gamma,\widebar{\sigma}), \\
& &&  \phi(\gamma,\sigma) > 0, \forall (\gamma, \sigma) \neq (r,c). 
\end{aligned}
\label{eq:error minimization}
\end{equation}
In this optimization problem, the first two constraints model the requirement that the matrix $\phi$ is a potential function for the bimatrix game $A_1+E, C_2$. The final constraint models the requirement that the global minimum of $\phi$ is exactly at the specified choice location $(r, c)$.

Notice the strict inequality for $\phi(\gamma,\sigma) > 0$. This manifests within $\phi$ as a small offset $\epsilon <10^{-6}$ added to entries that would be set to zero due to the exact potential constraints. This can be interpreted as a slack term that allows for a small deviation in order to enforce the global minimum.

Now our approach, summarized in Algorithm~\ref{alg:policy select}, computes the optimal error matrices for every choice of candidate global minima $(r,c)$ in $\phi$. Note that should the algorithm fail to find a suitable $E$ using any $(r,c)$, player 1 will default back to the scalarization method.

\begin{algorithm}
\caption{Policy Selection from Adjusted Costs}
\label{alg:policy select}
\begin{algorithmic}[1]
    
\State \textbf{Input:} Cost matrices $A_1, B_1, C_2$, Weights $\theta_1,\theta_2$ 
\State \textbf{Output:} Policies $\tilde{\gamma}^s, {\sigma}^s$ for players 1 and 2.
\State Compute security policy: ${\sigma}^s = \argmin \max C_2$
\State Initialize best error matrix: $E^* = \infty$
\For{$r$ in $\mathcal{M}_1(A_1,B_1,{\sigma}^s)$}
    \State Find optimal $(E, \phi)$ via Equation~\eqref{eq:error minimization} using $(r, {\sigma}^s)$.
    \If{$E$ is finite \textbf{and} $\norm{E}^2_F < \norm{E^*}^2_F$}
        \State $E^* = E$
    \EndIf
\EndFor
\State Compute,\[ 
\tilde{\gamma}^s = \begin{cases} \arg\min \max ({A}_1 + E^*), &\text{ if $E^*$ is finite}, \\
\arg\min \max (\theta_1A_1+\theta_2B_1), &\text{ otherwise.}\end{cases}
\]
\end{algorithmic}
\end{algorithm}

The following is a key property of Algorithm~\ref{alg:policy select}.

\begin{thm}[Output of Algorithm~\ref{alg:policy select}]
Given input matrices $A_1$ and $C_2$ and for any fixed set of weights $\theta_1, \theta_2$, the resulting policies $\{\tilde{\gamma}^s$, $\sigma^s\}$ form a pair of security policies for the bimatrix game $\tilde{A}_1, C_2$ and also a Nash equilibrium. 
\end{thm}

\textbf{Proof.} Note that in general bimatrix games, Nash equilibria are not equivalent to security policies. In fact, pure Nash equilibria are not guaranteed to exist in general bimatrix games. However, in the present case, we leverage the fact that all global minima of a bimatrix potential game correspond to Nash equilibria of the bimatrix game (\cite{NoncooperativeGame}). The optimization in Equation~\eqref{eq:error minimization} finds $E$ that defines $\tilde{A}_1$ such that there is a unique global minimum. The column $\sigma^s$ corresponding to the global minimum is selected to match the security policy $\sigma^s$ (line 6 of Algorithm~\ref{alg:policy select}). Finally, in line 13, we pick $\tilde{\gamma^s}$ to be the security policy corresponding to the best error matrix found, corresponding to the $\sigma^s$-th column. Therefore the Nash equilibrium policies are also equivalent to security policies $\{\tilde{\gamma^s},\sigma^s\}$.  \qed

\begin{remark}[{\bf Vector Cost Players}]
    In the case that two vector cost agents are playing against each other, Algorithm~\ref{alg:policy select} would be modified so that both players produce error matrices $E_1, E_2$ to force the global minimum at any entry. The policies produced would be $\gamma^s \in \mathcal{M}_1(A_1,B_1,\sigma^s)$ and $\sigma^s \in \mathcal{M}_2(A_2,B_2,\gamma^s)$. This would behave like a central planning problem since the optimization is much less constrained and the costs can be manipulated to produce any outcome that is desired.
\end{remark}

These policies provide an associated upper bound in cost error compared to the original outcome in the unadjusted vector game. We define two security policies with respect to $A_1$ and $B_1$ denoted as $\gamma^s_A$ and $\gamma^s_B$, both of whom lie within $\mathcal{P}_1(A_1,B_1,\sigma^s)$ and can be considered corner points on the Pareto frontier. These serve as the policies we will compare against the new policies resulting from the cost adjustment method. Assume that the method described in Equation~\eqref{eq:error minimization} was performed successfully for a given minimum entry at $(\tilde{\gamma}^s,\sigma^s)$ for bimatrix pair $A_1,C_2$, the output error matrix we name $E_A$. Now perform the same optimization with costs $B_1,C_2$ to produce error matrix $E_B$. With these error matrices defined, we can now describe the guarantee on the upper bound in outcome performance reduction for each cost type relative to the security policies $\gamma^s_A$ and $\gamma^s_B$ for using the cost adjustment algorithm, described below as:
\begin{multline*}
\begin{bmatrix}
        A_1(\tilde{\gamma}^s, \sigma^s) - A_1(\gamma_A^s, \sigma^s) \\
        B_1(\tilde{\gamma}^s, \sigma^s) - B_1(\gamma_B^s, \sigma^s)
\end{bmatrix}
    \leq 
    \begin{bmatrix}
        \norm{E_A}_F \\
        \norm{E_B}_F
    \end{bmatrix}\\
 \forall \tilde{\gamma}^s \in U_1, \forall \sigma^s \in U_2,
\label{max error}\end{multline*}
where the maximum departure from the original outcome value is never more than the norm of the error matrix for that cost type. Note that here we can take the error matrix as a quantification of how far the original priority set has been distorted in order to produce new costs that yield the desired security policy $\tilde{\gamma}^s$. This error can be thought of as the control effort necessary to force solutions into a desired set.

\begin{remark}[{\bf Error Measurement}]
    

   The goal of the optimization is to create the new cost matrix with the desired properties while minimizing the disturbance to the original costs. When comparing several error matrices, we select the one that causes the least deviation in a similar manner. In general it is difficult to distinguish which outcomes on the Pareto frontier are more desirable than others since it is a comparison between completely different quantities related to multiple distinct objectives. Here we use $||E||$ as a metric to pick the most feasible Pareto optimal solution in the sense of the original unadjusted costs. Based on this criterion, there can be a unique solution to the vector cost game with respect to a clear benchmark.
\end{remark}

While Theorem 1 addresses how to compute the preferred policies, it does not explicitly address the question of finiteness of the error matrix $E$. The next result characterizes the properties of the cost matrices $A_1, C_2$ such that Algorithm~\ref{alg:policy select} produces finite solutions $E^*$ and $\phi$.

To characterize the conditions on starting costs that yield solutions from Algorithm~\ref{alg:policy select}, we introduce a pairwise difference operation. We define pairwise column and row differences $D^{col}(\cdot) \in \mathbb{R}^{f \times m}$, $D^{row}(\cdot) \in \mathbb{R}^{n \times g}$, where $f:=({n \atop 2})$ and $g:=({m \atop 2})$, as matrices containing every combination of differences between elements in the argument matrices along each column and row, respectively. The relationship between entries in each column $j$ of a matrix $X$ can be defined as
\begin{equation*}
    D_{j}^{col}(X_j)=[x_{ij}-x_{kj}\,\,  | \,\, \forall i, k \ \; 1 \leq i < k \leq n].
\label{column difference}\end{equation*}
Likewise the row pairwise difference is defined across each row $i$ as
\begin{equation*}
    D_{i}^{row}(X_i)=[x_{ij}-x_{ik} | \forall j, k \; \; 1 \leq j < k \leq m].
\label{row difference}\end{equation*}

With these definitions, we now describe the constraints of the exact potential game from Equation~\eqref{eq:exact potential} more succinctly as
\begin{equation*}
    \begin{bmatrix}
        D^{col}(\phi) \\
        D^{row}(\phi)'
    \end{bmatrix}
    =
    \begin{bmatrix}
        D^{col}(A_1+E) \\
        D^{row}(C_2)'
    \end{bmatrix}.
\label{pairwise potential}\end{equation*}

The following result characterizes the set of rows and columns which can correspond to the global minimum.

\begin{thm}\label{thm:cost conditions}
For a given value of the costs $C_2$ for player 2, Equation~\ref{eq:error minimization} produces finite $E$ such that bimatrix game $\tilde{A}_1, C_2$ produces potential $\phi$ with a global minimum at chosen position $(r,c)$ if fixed costs $C_2$ adhere to the following conditions. For entry $d_{rj} \in D^{row}(C_2)$:
\begin{equation}\label{eq:cost conditions}
    \begin{aligned}
    & d_{rj}<0, \forall j < c, \quad d_{rj}>0, \forall j > c.
    \end{aligned}
\end{equation}
\end{thm}

\textbf{Proof.} The potential game can be expressed only in terms of the fixed costs $C_2$ that are unaffected by the cost adjustment algorithm and determine the viability of certain selections of global minimum. The potential game constraints are now simply $D^{row}(\phi)=D^{row}(C_2)$, with $D^{col}(A_1+E)$ as a free variable that can be adjusted to accommodate the pairwise row differences. When selecting a global minimum based on player 2's costs, the entries in the potential function must be of the form:
\begin{equation*}
    \phi = \begin{bmatrix}
        k_1-d_{11} & \ldots & k_1 & \ldots & k_1+d_{1g} \\
                \vdots &  &\vdots & &\vdots \\
        -d_{r1} & \ldots & 0 & \ldots & d_{rg} \\
        \vdots &  &\vdots & &\vdots \\
        k_n-d_{n1} & \ldots & k_n & \ldots & k_n+d_{ng}
    \end{bmatrix}.
\label{row differencce potential}\end{equation*}
Here every entry is defined in terms of the pairwise difference of the row of the original costs and arbitrary offset constant $k_i>0$ for each $i$th row. This constant is selected so that $\phi(r,c)=0$ is the lowest value, which serves to enforce that minimum across all rows.

Notice that for all rows other than row $r$, we can select a $k_i$ such that all of the row entries are greater than zero. For row $r$ however, entries $d_{rj}$ must be already be in a range that results in positive values across every entry that is not the minimum. If $d_{rj}>0$ for entries to the left of column $c$, the value will be negative and invalidate this constraint. Similarly, this will be the case if $d_{rj}<0$ for values to the right of column $c$. Therefore, in order to select global minimum $(r,c)$, the row pairwise differences of $C_2$ must adhere to the constraints described in Theorem~\ref{thm:cost conditions}. \qed

\begin{remark}[{\bf Fixed Cost Conditions}]
    In order for the optimization to converge successfully, there are $\ceil{\frac{1}{2} \times g}$ total unique conditions of the type described in Theorem~\ref{thm:cost conditions} that must be met by the original fixed costs. Checking that a simple majority of the entries within a row satisfy the constraints will determine whether the minimum is viable. Because of this, we can determine a priori from the costs of player 2 whether the algorithm will converge. If not, either another minimum must be selected or the starting costs must be redesigned. \sdbcheck{The computational complexity of Algorithm~\ref{alg:policy select} is measured in practice to be approximately $\mathcal{O}(n^3)$ with square matrices of size $n=m$ for the optimized implementation using CVX. This is suitable for considering small action spaces in real-time applications.}
\end{remark}

\section{Numerical Comparison with Baseline}\label{sec:numerics}

\begin{figure}
    \centering
    \includegraphics[width=0.5 \linewidth]{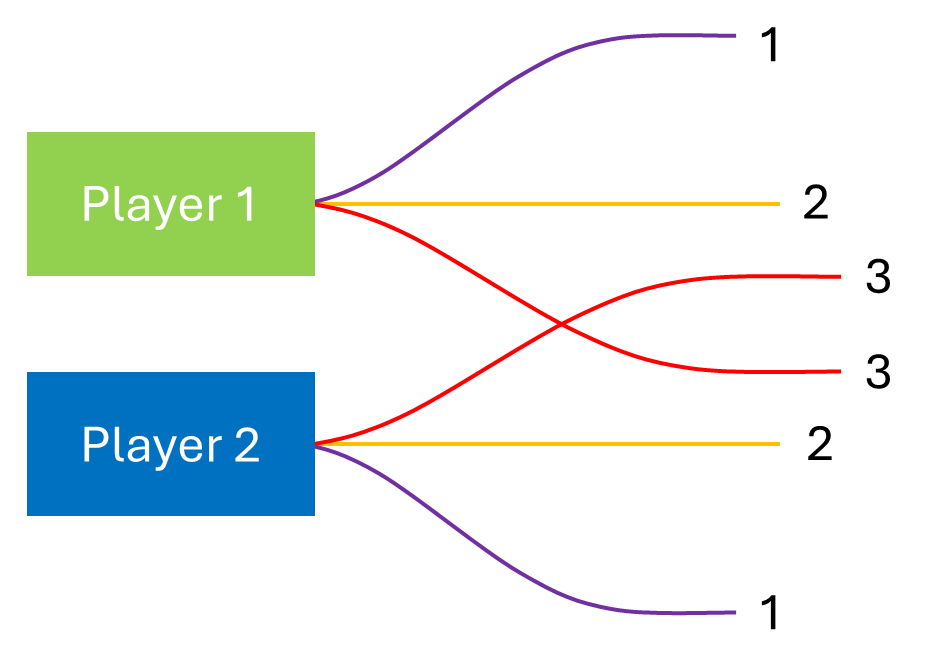}
    \caption{\small Example Trajectory Planning Scenario}
    \label{fig:trajectory}
\end{figure}

Here we will outline a simple example involving two vehicles with some overlapping possible trajectories, pictured in Figure~\eqref{fig:trajectory}. Trajectories one through three for player 1 correspond to rows one through three in the matrices of Equation~\eqref{eq:base game}, likewise for player 2 across each column. The base vector cost game can be taken as two bimatrix games being played simultaneously, as in  
\begin{equation}
    \begin{aligned}\label{eq:base game}
        A_1 = 
        \begin{bmatrix}
            0 & 1 & 2 \\
            -1 & 0 & 1\\
            -2 & -1 & 0
        \end{bmatrix}, 
         B_1 = 
        \begin{bmatrix}
            0 & 1 & 2 \\
            1 & 2 & 3\\
            2 & 3 & 4
        \end{bmatrix},
        A_2 = -A_1, B_2 = B_1 
    \end{aligned}
\end{equation}

The logical choice independent of the cost structure would be for both players to choose trajectory 2, which minimizes competitive costs as much as possible while avoiding the possibility of a collision, which would result in the highest safety cost. This preference is captured using the set of Pareto policies $\gamma=2 \in \mathcal{P}_1(A_1,B_1,2)$, and the outcome of this policy set would be $J_i(2,2)=(0,2)$. 
Observe that there is no particular reason to pick $\gamma=2$ based on Pareto optimality alone since any outcome is technically Pareto optimal for a fixed $\sigma$. With a more complex situation, selecting the desired policy to balance each objective would become quite ambiguous. We could achieve solutions at either extreme of the Pareto frontier simply by playing security policy $\gamma=3$ to minimize cost $A_1$ or $\gamma=1$ to minimize cost $B_1$. However, if we desire policies that avoid extreme outcomes for any of the cost types, we need to add additional structure to the problem.

Scalarization can give more control over the preferences such as this, allowing us to weight some costs more heavily and change the behavior of the players. Selecting $\theta_1=2, \theta_2=1$, the combination of costs according to Equation~\eqref{scalarization} results in
\begin{equation*}
    C_1 =
    \begin{bmatrix}
        0 & 3 & 6 \\
        -1 & 2 & 5\\
        -2 & 1 & 4
    \end{bmatrix}, 
        \quad C_2 = C_1^T.
\end{equation*}
This cost setup prioritizes competitive costs at the expense of safety, making $\{\gamma^s,\sigma^s\}=\{3,3\}$. \sdbcheck{The outcome of this exchange would be $J_i(3,3)=(0,4)$ for the original costs, a Nash equilibria within the space of the scalarized costs. From a practical perspective this is not an ideal outcome despite the equilibrium, since this cost structure incentivizes collisions.}

Unfortunately in this case we cannot pick a pair $(\theta_1,\theta_2)$ that results in an equilibria at $(2,2)$. Reducing the weight of $\theta_1$ to intermediate values does not change the security policy outcome as long as $\theta_1 > \theta_2$. Security policies will always be Nash equilibria within a single game type, but result in worst-case outcomes in the other game. 

As an alternative to settling for suboptimal scalarization or completely revising the underlying cost structure, we apply our convex optimization to force the costs to admit the desired security policy $\tilde{\gamma}^s=2$. We assume that only player 1 is using this technique and that player 2 plays a security policy over scalarized costs with the weights $\theta_1=2, \theta_2=1$. Even with fixed costs for player 2, the convex optimization finds the matrix $E$ for which $\tilde{A}_1$ has $\tilde{\gamma}^s=2$.
\begin{equation*}
    \begin{aligned}
        E = \begin{bmatrix}
             0 & 0 & 0 \\
            -0.5 & -0.5 & -0.5\\
            0.5+\epsilon & 0.5+\epsilon & 0.5+\epsilon
        \end{bmatrix},
        \quad \phi = 
        \begin{bmatrix}
             3.5 & 2.5 & 1.5 \\
             2 & 1 & 0\\
             2+\epsilon & 1+\epsilon & \epsilon
        \end{bmatrix}
    \end{aligned}
\end{equation*}
Examining $\phi$ shows a global minimum corresponding to $\{\tilde{\gamma}^s,\sigma^s\}=\{2,3\}$ with a small $\epsilon$ on the order of $10^{-6}$ supporting this as the lowest value in the matrix. Player 1's policy is $\tilde{\gamma}^s \in \mathcal{M}_1(A_1, B_1,\sigma^s)$.

The tolerance $\epsilon$ captures the amount of slack given to the differences in potential across each column. Player 1's policy is $\tilde{\gamma}^s \in \mathcal{M}_1(A_1, B_1,\sigma^s)$. If the cost structure of $C_2$ adheres to the constraints in Equation~\eqref{eq:cost conditions} for any given coordinate of global minimum, we can adjust the costs of $A_1$ to create a minimum with matching security policies at that point. Since we are not changing the entries of $C_2$, all viable global minima for the cost adjustment lie along the column matching $\sigma^s=3$. 

\section{Application to autonomous racing}\label{sec:application}

In this section, we explore the application of the vector cost game to the task of autonomous vehicle racing. We compare the performance of our approach against a scalarization approach in a passing interaction, where one vehicle, designated as the attacker, tries to pass the slower defending vehicle in front of it. The different cost structures of each player type result in behavior that favors particular objectives at the expense of others, the details of which are discussed below.

The dynamics for the vehicles used in the racing simulation were implemented according to the kinematic bicycle model (\cite{PotentialGameBased}) for each player $i$ with a time step of $\Delta t$. State variables include $x_i$ and $y_i$ for position, $v_i$ for velocity, $\psi_i$ for heading, and \sdbcheck{$\beta_i$ for simplified sideslip}. The control inputs are $a_i$ for acceleration and $\delta_i$ for steering angle. The constants $l_r$ and $l_f$ are determined from the lengths of the vehicle along the $x$ and $y$ axes, respectively (\cite{PotentialGameBased}). The relationships between each of these model parameters are given as

\begin{equation*}\label{eq:bicycle dynamics}
    \begin{aligned}
        x_i(t+1) &= x_i(t) + v_i(t) \cos(\psi_i(t) + \beta_i(t)) \Delta t, \\
        y_i(t+1) &= y_i(t) + v_i(t) \sin(\psi_i(t) + \beta_i(t)) \Delta t, \\
        v_i(t+1) &= v_i(t) + a_i(t) \Delta t, \\
        \psi_i(t+1) &= \psi(t) + \frac{v_i(t)}{l_r} \sin(\beta_i(t)) \Delta t, \\
        \beta_i(t+1) &= \tan^{-1} \left( \frac{l_r}{l_r + l_f} \tan(\delta_i(t)) \right) \Delta t.
    \end{aligned}
\end{equation*}

\begin{figure}
    \centering
\includegraphics[width=0.95\linewidth]{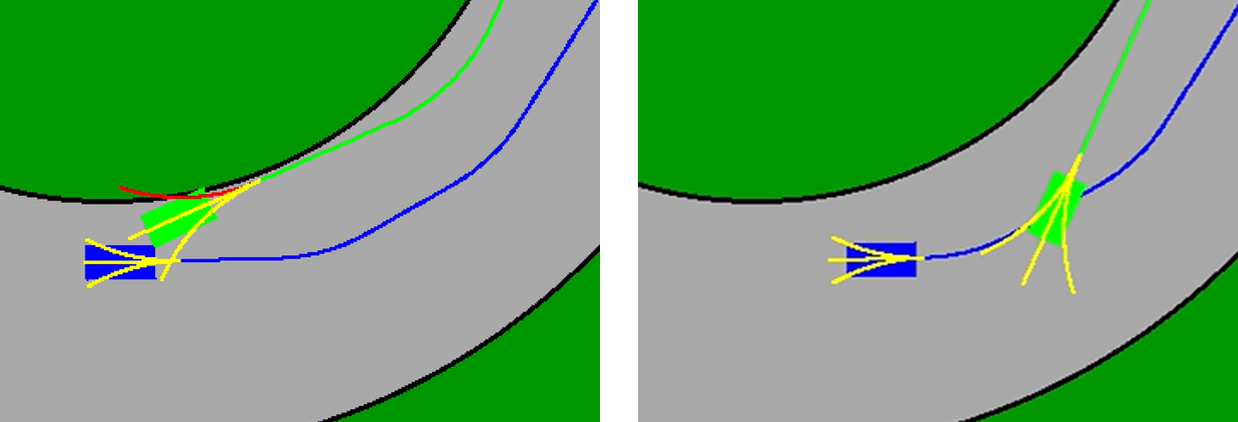}
    \caption{\small Scenario I (left): risky passing maneuver, resulting in a collision. Scenario II (right): safe passing maneuver.}
    \label{fig:race scenarios}
\end{figure}

The action space used in the bimatrix game has $n=m=9$ total choices. Each choice is a trajectory created by a combination of acceleration and steering angle, which are held constant across the simulation horizon until the next decision is made. These correspond to each combination of speeding up, slowing down, and maintaining speed for the acceleration input and turning left, turning right, and going straight for the steering angle input. We designed a racing simulator to visualize the results using Pygame, overlaid with several custom Python classes for the dynamics, opponent sensing, and optimization algorithm. Visuals for Scenarios I and II can be seen in Figure~\ref{fig:race scenarios}. 

The cost structure is identical to that described in Section~\ref{sec:problem}. Costs $A_i$ are generated by comparing the relative distance between the terminal point on each trajectory of the action space to that of the opponent, rewarding progress at the other player's expense. \sdbcheck{Costs $B_i$ contain penalties for undesirable outcomes. Each point that leaves the track incurs a small penalty which is summed over the entire trajectory, while intersecting trajectories with those of the opponent receive a large one-time cost.} For the scalarization of costs, \sdbcheck{we tuned the values of $\theta_1$ and $\theta_2$ heuristically until the observed behavior of player 2: (i) consistently stayed on the track and, (ii) avoided collisions whenever possible.}

Algorithm~\ref{alg:policy select} was applied during each decision making epoch, assuming that player 2 had already selected $\sigma^s$. This is  similar to the Stackelberg game (\cite{ANoncooperativeGame}). For this particular cost structure, the cost adjustment process was successful in finding a viable global minimum for the convex optimization 49.15$\%$ of the time, otherwise defaulting to the scalarization approach.

For each race, we solved a number of consecutive bimatrix games over a single horizon with an action space of nine trajectories for each player. We examined three different scenarios -- (I) Both players using scalar costs while attacking and defending, (II) Player 1 using Algorithm~\ref{alg:policy select} while attacking and player 2 defending using the scalarization approach, and (III) Player 1 using Algorithm~\ref{alg:policy select} while defending and player 2 attacking using the scalarization approach. Each player was spawned with a small initial velocity at a random position along the center line of the track. The attacking player was given a max speed 50\% higher than the defending player. The same sequence of start positions was repeated for 100 different races in each scenario.

\begin{table}[h]
    \centering
    \renewcommand{\arraystretch}{1.3}
    \caption{Race Event Frequency}
    \label{tab:race events}
    \begin{tabular}{cccc}
        \hline
        \makecell{\textbf{Scenario} \\ } & 
        \makecell{\textbf{Passes}} & 
        \makecell{\textbf{Collisions}} & 
        \makecell{\textbf{Attacker} \\ \textbf{Off Track}} \\
        \hline
        I & 38 & 55 & 11 \\
        II & 27 & 21 & 39 \\
        III & 31 & 47 & 33 \\
        \hline
    \end{tabular}
\end{table}

Referring to the the data in Table~\ref{tab:race events}, we see that in general player 2 raced much more aggressively than player 1. The specific combination of $\theta$ values tended to create higher competitive costs relative to safety costs, which subsumed the goals of avoiding collisions and going off the track to an extent. Looking at the performance of player 1, we see that sacrificing a small number of passing attempts allowed both players to avoid collisions much more effectively. 

Note that while there is a high count of off the track instances for player 1 while in the attack role, there are a number of mitigating factors. Many of these occurred during the middle of passing maneuvers when the player had to choose directly between colliding with the opponent and leaving the track. Since out of bounds penalties were set to be lower than collision penalties by design, this behavior is not entirely undesirable. For the purpose of our analysis, we focused on only two costs, which causes all concerns other than competitive goals to be condensed into a single cost $B_i$. Splitting collision costs and out of bounds costs into separate matrices and performing the optimization over a Pareto frontier with three separate costs would likely remedy this.

\begin{table}[h]
    \centering
    \renewcommand{\arraystretch}{1.3}
    \caption{Attacker and Defender Progress}
    \label{tab:race progress}
    \begin{tabular}{cccc}
        \hline
        \makecell{\textbf{Scenario} \\ } & 
        \makecell{\textbf{Attacker Lead} \\ \textbf{Time (\%)}} & 
        \makecell{\textbf{Attacker} \\ \textbf{Avg. Laps}} & 
        \makecell{\textbf{Defender} \\ \textbf{Avg. Laps}} \\
        \hline
        I & 64 & 1.61 & 0.88 \\
        II & 58 & 1.51 & 0.90 \\
        III & 62 & 1.55 & 1.05 \\
        \hline
    \end{tabular}
\end{table}

Table~\ref{tab:race progress} shows statistics for how well the attacker did in each scenario. A successful attacker can be said to maintain the lead for the longest amount of time, maximize their progress around the track, and minimize the progress of the defender when possible. Player 1 experienced a slight decrease in attack performance using the cost adjustment approach compared to scalar costs, but only sacrificed 6\% lead time and 0.1 laps worth of progress. This is a small price to pay for a 62\% decrease in collision rate. While defending, player 1 performed better than player 2 in Scenario III since player 2 reacted poorly to some of the vector cost choices made by player 1, resulting in more off the track incidents. 

\section{Conclusion and Future Directions}\label{sec:conclusion}

In this paper, we presented a vector cost structure for bimatrix games as an alternative to scalarization. \sdbcheck{It provides the designer with additional control over how multi-objective problems are solved and yields outcomes with desirable properties such as Pareto optimality, guarantees with respect to each objective, and pure Nash equilibrium solutions.} The cost adjustment approach was successfully applied to an autonomous racing scenario between two players, one using scalarization and the other using the vector cost convex optimization algorithm. Our approach resulted in a dramatic decrease in collision rate with a small drop in passing performance. 

Extensions of this approach could include the applications to more than two costs or more than two players. It could be used to guarantee minimum performance for additional objectives in any scenario where cost design is needed. 

\renewcommand{\refname}{}  
\section{References}\label{sec:references}
\vspace{-1.5em}  
{\small \bibliography{main}}

\end{document}